\documentclass[runningheads]{llncs}
\usepackage[T1]{fontenc}

\usepackage{graphicx}
\usepackage{orcidlink}
\usepackage{amsmath}
\usepackage{amssymb} 
\usepackage{booktabs} 
\usepackage{array} 
\usepackage{graphicx} 
\usepackage{multirow} 
\usepackage{caption} 
\usepackage{subcaption}
\usepackage{float}  
\usepackage{placeins}  
\newcommand{\etal}{\textit{et al.}}
\usepackage{soul}
\usepackage{xcolor}

\begin{document}
\sloppy
\raggedbottom
\title{Cross Feature Fusion of Fundus Image and Generated Lesion Map for Referable Diabetic Retinopathy Classification}
\titlerunning{Cross Feature Fusion for DR Classification}
%
\author{Dahyun Mok\inst{1} \and
Junghyun Bum\inst{2}\orcidlink{0000-0002-9926-2910} \and
Le Duc Tai\inst{1}\orcidlink{0000-0002-5286-6629} \and
Hyunseung Choo\inst{1,*}\orcidlink{0000-0002-6485-3155}}
\authorrunning{D.Mok et al.}

\institute{Dept. of Electrical and Computer Engineering, Sungkyunkwan University, Korea \and
College of Computing and Informatics, Sungkyunkwan University, Korea \\
*Corresponding author \\
\email{\{dahyun1025,bumjh,ldtai,choo\}@skku.edu}}

\maketitle 

\begin{abstract}
Diabetic Retinopathy (DR) is a primary cause of blindness, necessitating early detection and diagnosis. 
This paper focuses on referable DR classification to enhance the applicability of the proposed method in clinical practice. 
We develop an advanced cross-learning DR classification method leveraging transfer learning and cross-attention mechanisms. 
The proposed method employs the Swin U-Net architecture to segment lesion maps from DR fundus images. 
The Swin U-Net segmentation model, enriched with DR lesion insights, is transferred to generate a lesion map.
Both the fundus image and its segmented lesion map are used as complementary inputs for the classification model. 
A cross-attention mechanism is deployed to improve the model's ability to capture fine-grained details from the input pairs. 
Our experiments, utilizing two public datasets, FGADR and EyePACS, demonstrate a superior accuracy of 94.6\%, surpassing current state-of-the-art methods by 4.4\%. 
To this end, we aim for the proposed method to be seamlessly integrated into clinical workflows, enhancing accuracy and efficiency in identifying referable DR.

\end{abstract}

\section{Introduction}

\begin{figure}[htbp]
  \centering
  \begin{subfigure}[t]{0.6\textwidth} 
    \centering
    \includegraphics[width=\textwidth]{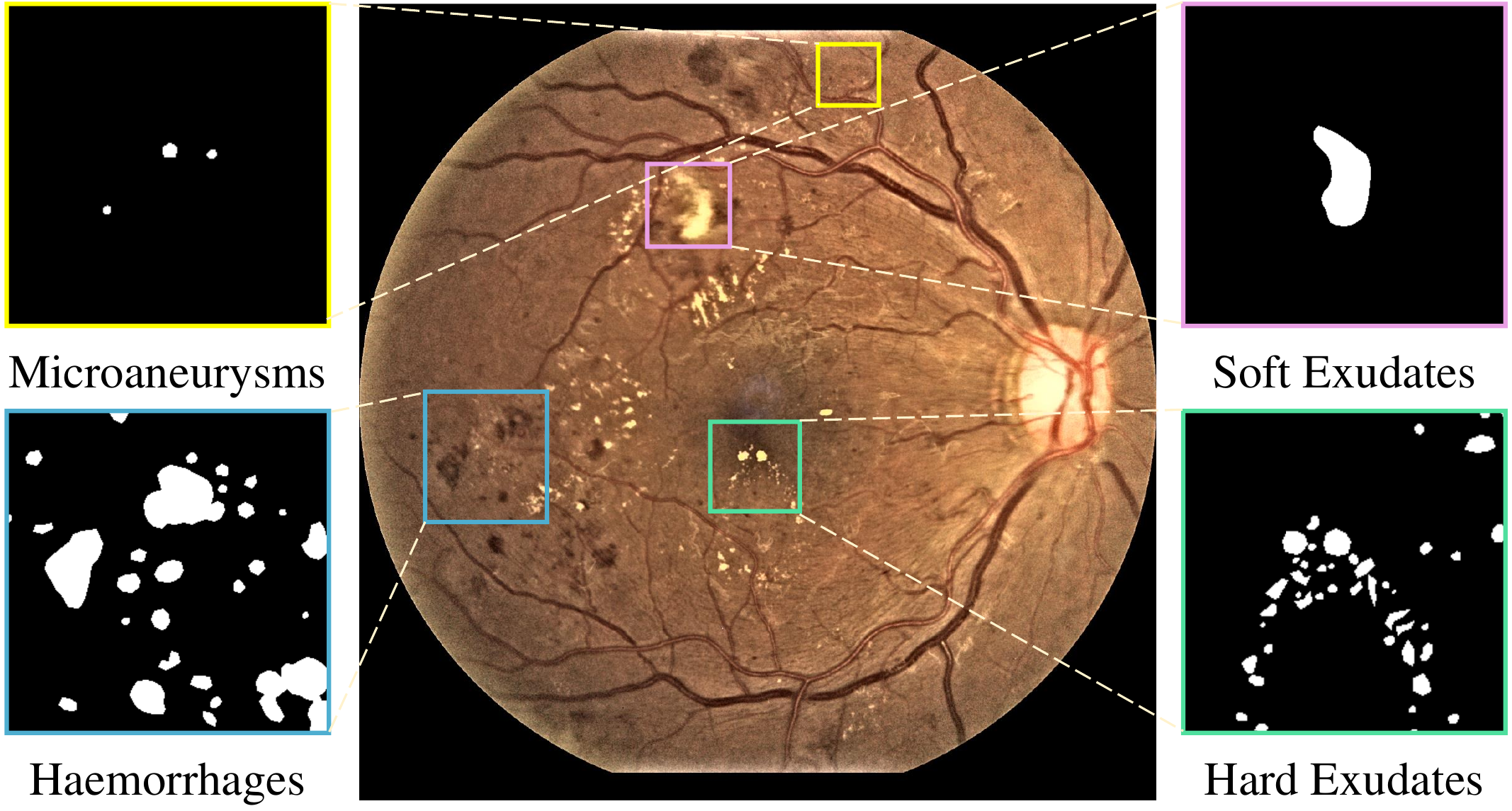} 
    \caption{}  
    \label{fig:example_1_a}
  \end{subfigure}
  \hfill
  \begin{subfigure}[t]{0.32\textwidth} 
    \centering
    \includegraphics[width=\textwidth]{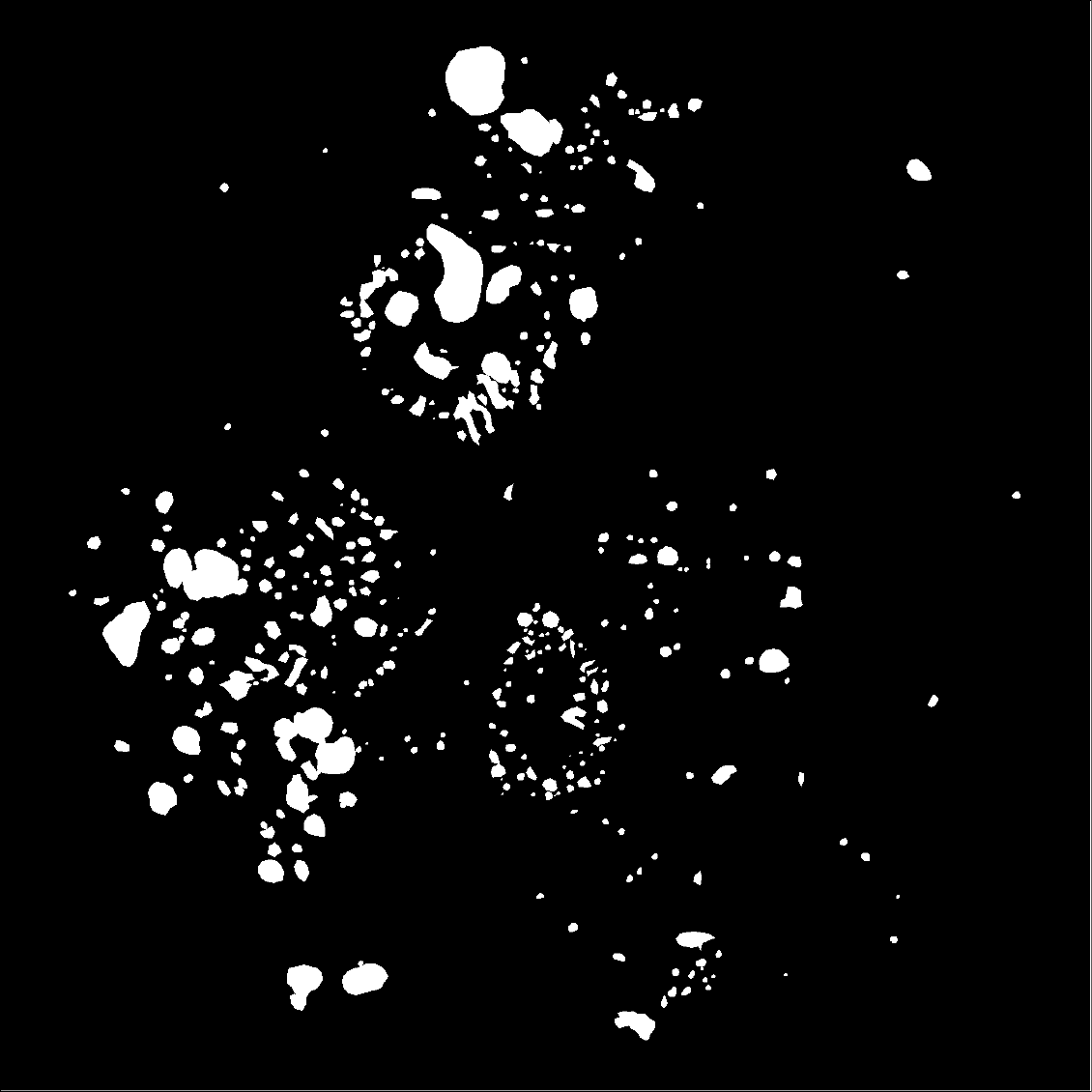} 
    \caption{}  
    \label{fig:example_1_b}
  \end{subfigure}
  
  \caption{Example of DR signs: (a) showing microaneurysms, hemorrhages, and exudates, clearly illustrating how each lesion appears in DR images and (b) showing the examples of lesion map.}
  \label{fig:1}
\end{figure}

\noindent Diabetic Retinopathy(DR) is a serious and prevalent complication of diabetes mellitus, posing a significant public health concern due to its potential to cause irreversible blindness if left untreated\cite{01}. 
DR develops as chronic hyperglycemia damages the small blood vessels within the retina. When these vessels are compromised, they can leak fluid or blood, leading to visual impairment. 
Without rapid treatment, DR can advance from mild, non-proliferative stages to severe proliferative stages, characterized by the growth of abnormal new blood vessels on the retinal surface, potentially causing significant vision loss or blindness\cite{03,02}. 
As depicted in Fig.~\ref{fig:1}, Microaneurysms, the earliest clinical signs of DR, manifest as small red dots caused by weakened capillary walls. 
Hemorrhages occur when these fragile vessels rupture, resulting in small blot or flame-shaped red spots. 
Exudates, which are lipid residues, appear as yellow or white spots on the retina, often forming a circinate pattern.
Proliferative DR is distinguished by neovascularization, where new and abnormal blood vessels form on the retinal surface, posing a high risk of severe vision impairment if they bleed into the vitreous humor.

Despite significant advancements in applying Convolutional Neural Network(CNN) for DR detection, achieving high accuracy in referable DR classification remains a challenging task. 
The subtle and heterogeneous characteristics of retinal lesions and the diversity of data collection environments can degrade the performance of deep learning networks, so integrating lesion-specific insights into the classification process can address these challenges more effectively\cite{04}. 
In medical applications, where diagnostic accuracy is paramount, inaccuracies can lead to delayed treatments or unnecessary interventions, impacting patient outcomes. 
While CNN has shown promise, its limitations necessitate the development of more robust methods. 
Therefore, this study proposes a novel approach that combines transfer learning and cross-attention mechanisms to improve the precision and reliability of DR classification. 
By incorporating lesion-specific insights into the classification process, our method overcomes the limitations of existing models and provides a more effective tool for early DR detection.

In this paper, we propose a cross-learning framework to enhance the classification of referable DR. 
Firstly, we introduce a DR segmentation method that uses the Swin U-Net architecture for segmenting lesion maps from retinal fundus images. 
Secondly, we generate lesion maps using the model trained in the first stage. Then, using the generated lesion maps, i.e. pseudo-lesion maps and the original images, we perform cross-learning to train the classification model. 
The integration of original images with pseudo-lesion maps via cross-fusion represents a novel approach in the medical imaging field.
Segmentation annotations for medical images are costly to obtain and typically limited to small datasets. Our method effectively handles cases where datasets are imbalanced, with abundant image-level labels but limited pixel-level segmentation labels, enhancing generalizability across datasets.
Our experimental results show an outstanding classification accuracy of 94.6\%, surpassing existing state-of-the-art methods. 
The cross-attention mechanism allows the model to focus on relevant regions of the image, enhancing its ability to distinguish different DR stages.

The remainder of this paper is structured as follows: In Section 2, we review related work on DR classification using deep learning and discuss recent technological advancements in this domain. 
Section 3 provides a detailed explanation of the model architecture and the proposed cross-learning framework. 
Section 4 presents data preprocessing steps, our experimental setup, and results, and offers an in-depth analysis of these findings, including both quantitative and qualitative assessments. 
Finally, Section 5 concludes the paper with a summary of our experiments and suggestions for future research directions. 

\section{Related Work}

\noindent DR classification has seen significant advancements with deep learning techniques. 
CNN has been widely employed for the automatic detection and classification of DR, demonstrating high accuracy and robustness in various studies. 
For example, ensemble models combining multiple CNN architectures have been shown to enhance performance by leveraging the strengths of individual models\cite{08,06,05}. 
Techniques such as transfer learning, where pre-trained models on large-scale datasets are fine-tuned for DR classification tasks, have significantly reduced the need for extensive labeled data\cite{07}. 
Additionally, integrating attention mechanisms has allowed models to focus on clinically relevant regions in retinal images, improving diagnostic accuracy. 
These approaches have collectively contributed to the development of reliable and efficient DR classification systems\cite{02}. 
Recent advancements have also seen the use of hybrid models that combine CNN with other architectures to improve performance. 
For instance, attention-based CNN has been employed to enhance the focus on relevant regions of retinal images, thereby improving the model's ability to detect and classify lesions accurately\cite{09}. 
Additionally, studies have explored the use of generative adversarial networks (GANs) to augment training data, addressing the issue of limited labeled datasets and further enhancing model performance\cite{10}.

Advanced deep learning models demonstrate exceptional performance in various medical image analysis tasks, enabling more effective detection and diagnosis of complex lesions. 
For example, the Swin Transformer effectively captures long-range dependencies, making it ideal for processing high-resolution medical images. 
Its hierarchical and shifted window mechanism excels in detailed lesion detection, proving particularly useful for retinal diseases and other complex lesions\cite{11}. 
The high-resolution Swin Transformer achieves high accuracy in medical image segmentation, especially in accurately segmenting images of tissues with complex structures such as accurately delineating the boundaries of neural tumors, thereby contributing to high-precision diagnostics and treatment planning\cite{12}. 
The U-Net Transformer combines self and cross-attention mechanisms, delivering outstanding performance in medical image segmentation. 
By maintaining the strengths of U-Net while integrating the benefits of Transformers, it enables more precise segmentation of complex organ structures\cite{13}. 
The MV-Swin-T incorporates multi-view learning, significantly improving mammogram classification by effectively combining information from various angles to enhance accuracy, playing a crucial role in the early detection of cancer\cite{14}. 
These advancements underscore the continuous evolution of DR detection technology towards more accurate and interpretable systems.

Traditional DR classification methods still face performance limitations despite significant advancements. 
CNN-based approaches are vulnerable to the fine details of early-stage DR lesions and the variability in retinal image quality\cite{15}. 
These methods may fail to detect lesions accurately in the early stages. 
Adversarial training proposed by S. Zhao \etal, improves model robustness but balancing it with accuracy poses a challenge for clinical implementation\cite{16,17}. 
However, it helps models become more resilient to various attacks. The interpretability of model decisions remains a critical issue. Many deep learning models operate as black boxes, making it difficult to understand the rationale behind their predictions\cite{18}. 
Ensemble models and multi-modal data integration have been proposed to address this. These techniques combine the predictions of various models to improve performance and integrate multiple data sources to provide richer information. 
However, stronger, more accurate, and interpretable solutions are still needed in DR detection\cite{19}. Approaches utilizing visual explanations enhance model interpretability, aiding medical professionals in trusting and understanding model predictions\cite{06}. 
These techniques clarify the features upon which the model bases its decisions, especially useful in DR detection. Lesion-aware contrastive learning proposed by S. Cheng \etal, has improved DR diagnosis accuracy by focusing on lesion-centric feature learning, effectively detecting fine lesions in the early stages\cite{20}. 
This method reduces the need for large annotated datasets and improves model generalizability. Despite these technical advancements, there are limitations to applying early DR detection in clinical practice. 
Therefore, we propose a novel method integrating cross-attention and Swin Transformer Tiny (Swin-T) to address these challenges. 
This approach enhances the accuracy and interpretability of DR classification while maintaining computational efficiency and robustness to image quality variability. 
By effectively combining the features of the original image and the generated lesion map using cross-attention, the proposed method maximizes DR classification performance.

\newpage
\section{Methodology}

\begin{figure} [t]
  \centering
  \includegraphics[width=0.95\textwidth]{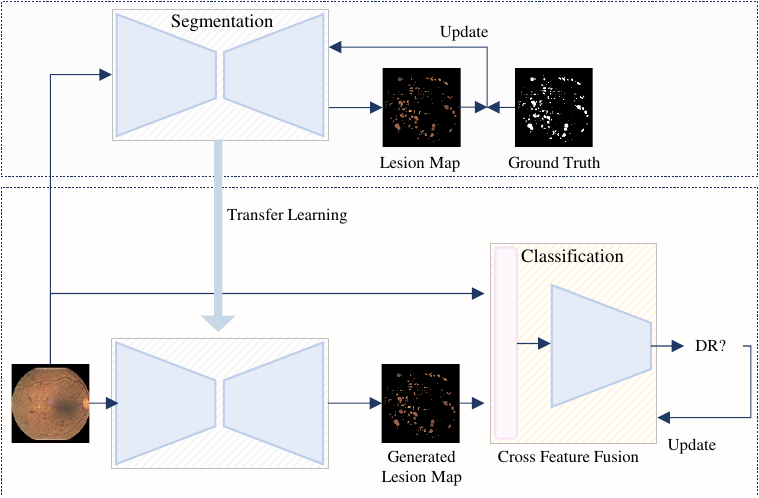}
  \caption{Overview of the proposed method consists of two steps.
  }
  \label{fig:2}
\end{figure}

\noindent \t Our proposed method improves referable DR classification performance through a two-step learning process: pixel-level lesions segmentation and image-level classification. 
Effective feature extraction is essential to enhance performance, and our proposed method sequentially utilizes the Swin U-Net and Swin-T models. 
As shown in Fig.~\ref{fig:2}, the first training process uses the Swin U-Net model to extract lesion maps from input fundus images. 
We use the trained model to generate lesion maps from the input fundus data. 
This domain adaptation step ensures the model leverages the lesion map features obtained from the first training step. 
Next, both the original fundus images and the lesion maps generated in the first step are fed into the Swin-T model pre-trained on ImageNet. 
Training steps combine the original fundus image embedding sequences with the lesion map embedding sequences through a cross-attention block. 
The cross-attention mechanism enhances the model's ability to focus on highly relevant regions, improving the classification accuracy for referable DR. 
As part of the training process, the "Update" step indicates the model parameter updates during backpropagation, which is critical for optimizing model performance.
This proposed two-stage learning method integrates lesion map segmentation, transfer learning, and cross-attention to provide a robust and accurate classification system.

\begin{figure}[H]
  \centering
  \includegraphics[width=0.95\textwidth]{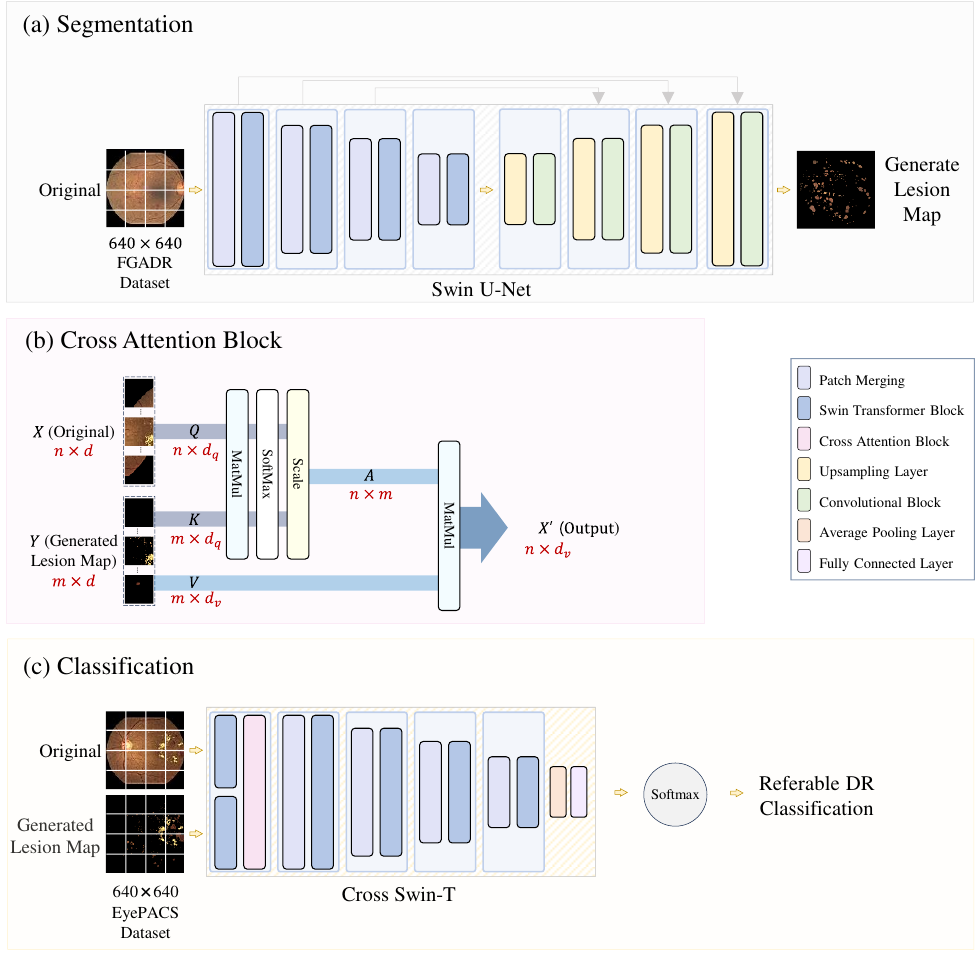}
  \caption{Architecture of the proposed method: (a) using the transfer learning capabilities of Swin U-Net, generates the lesion map. (b) illustrates a Cross Attention Block where original image \textit{X} and the generated lesion map \textit{Y} are concatenated and transformed into Query \textit{Q}, Key \textit{K}, and Value \textit{V} matrices to compute the final attended representation for \textit{X}. In (c), the original image and the generated lesion map are input into Cross Swin-T to classify DR.
  } 
  \label{fig:3}
\end{figure}

\subsection{Model Architecture}
\noindent In Step 1, the Swin U-Net architecture is utilized for DR lesion segmentation. 
The Swin Transformer, known for its hierarchical and shifted window attention mechanisms, serves as the encoder, capturing rich contextual information. 
The encoder consists of multiple stages, each with several Swin Transformer blocks that process image patches using self-attention mechanisms, capturing long-range dependencies and fine-grained details essential for accurate segmentation. 
The encoder also includes patch merging, which combines smaller image patches into larger ones to reduce the spatial dimensions while preserving essential features. 
The decoder, composed of upsampling layers and convolutional blocks, reconstructs the segmented lesion map from the encoded features, with skip connections ensuring effective use of both low- and high-level features. 
The segmentation model is trained using binary cross entropy with logits loss, ensuring it distinguishes between lesion and non-lesion pixels effectively, as shown in the following equation, where \( N \) is the number of samples, \( i \) is the sample index, \( y_i \) is the true label of the \( i \)-th sample, and \( p_i \) is the predicted probability for the \( i \)-th sample:

\begin{equation}
\text{BCE with Logits Loss} = \frac{1}{N} \sum_{i=1}^N \left[ \max(x_i, 0) - x_i y_i + \log \left(1 + e^{-|x_i|}\right) \right]
\end{equation}

For the classification task, the Swin-T architecture is utilized.
This model processes the input fundus images and their respective lesion maps using a cross-attention mechanism, capturing the interactions between the original images and the lesion maps to enhance the model's ability to differentiate between subtle features.
The cross-attention mechanism allows the model to integrate information from both the original image and the lesion map, focusing on critical regions for accurate classification and detect DR lesions.
The classification model is trained using binary cross entropy with logits loss, which helps the model output logits that are converted into probabilities during training, enhancing its ability to predict the probability of DR stages accurately.
At the final stage, a feature map is produced and subsequently passed through a fully connected layer to calculate the probability of DR.

\subsection{Proposed Method}
\noindent Transfer learning and a cross-attention mechanism are applied to enhance model performance.
Fig.~\ref{fig:3} shows the architecture of the proposed method. 
The encoder of the Swin U-Net segmentation model, pre-trained on the FGADR dataset, is used to input the EyePACS dataset's original images to obtain lesion maps. 
This retains lesion-specific features learned during segmentation, providing a rich feature set for the classification task. 
The cross-attention mechanism integrates information from both the original fundus image and the generated lesion map, i.e. pseudo-lesion map, improving focus on relevant regions. 
This dual approach ensures the model leverages global and local features, enhancing diagnostic accuracy. By incorporating cross-attention, the model dynamically prioritizes regions of interest in both input modalities, resulting in more precise and reliable DR classification.

In Step 1, the network is trained to segment lesion maps from retinal fundus images using the FGADR dataset. The Swin U-Net initially utilizes pre-trained weights from the ImageNet dataset for general visual knowledge. 
Subsequently, the model is fine-tuned on the FGADR dataset to learn specific lesion segmentation. The trained model generates a lesion map of the EyePACS data during the classification task. 
This two-step transfer learning process captures both general and domain-specific features, enhancing segmentation accuracy. 
Fine-tuning involves adjusting model weights to better fit the FGADR dataset, ensuring optimal performance in segmenting DR lesions.

Step 2 includes two end-to-end phases: generating lesion maps from the pre-trained model and classifying DR using a cross-attention mechanism.
After training the segmentation model in Step 1, the model is transferred to the classification model. 

By utilizing the pre-trained model, the classification model benefits from rich lesion-specific features, enhancing its ability to differentiate between various stages of DR. 
The model's input consists of the original fundus images and the lesion maps generated by the pre-trained Swin U-Net, and it outputs predictions regarding the presence of DR.
The input tensors \( X \) (original image) and \( Y \) (lesion map) are transformed into Query \( Q \), Key \( K \), and Value \( V \) as follows:
\begin{equation}
Q = X W_Q, \quad K = Y W_K, \quad V = Y W_V
\end{equation}
The attention weights \(A\) are calculated as:
\begin{equation}
A = \text{SoftMax}\left(\frac{Q \cdot K^T}{\sqrt{d_k}}\right)
\end{equation}
Unlike multi-head attention, which focuses on self-attention, our method's cross-attention mechanism effectively differentiates foreground and background by learning the interaction between the two inputs. This enhances the focus on critical lesion areas, improving the model’s performance.
Consequently, this approach significantly improved classification accuracy, demonstrating excellent performance in detecting subtle lesions and accurately distinguishing different stages of DR. 
The cross-attention mechanism further enhances this process by allowing the model to focus on the most relevant regions of the input images, ensuring accurate classification.
This stage includes fine-tuning the classification model to optimize performance. 
Dropout layers are incorporated to prevent overfitting and improve generalization by randomly deactivating some neurons during training. 
This addresses issues from diverse and complex data\cite{01}. 
The learning rate of the stochastic gradient descent optimizer diminishes over time, and binary cross entropy is employed for classification. Fine-tuning ensures the model achieves optimal performance, balancing accuracy and generalization.
This stage entails meticulous adjustment of hyperparameters to ascertain the model's robustness and efficacy in clinical settings.

\section{Experiments and Results}
\subsection{Datasets and Experimental Setup}

\begin{figure}[t]
  \centering
  \begin{subfigure}[t]{0.45\textwidth}
    \centering
    \includegraphics[width=\textwidth]{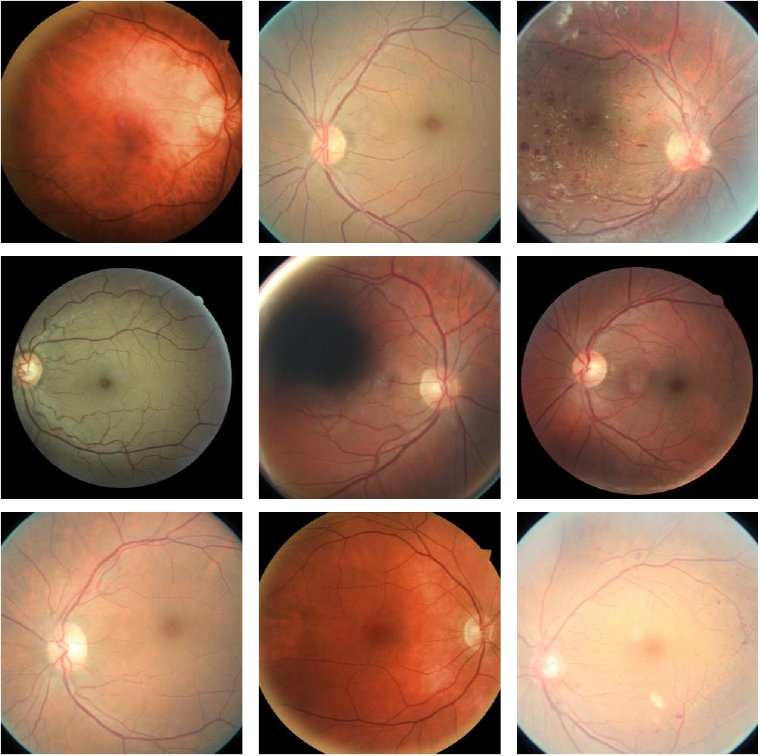}
    \caption{}
    \label{fig:example_4_a}
  \end{subfigure}
  \hspace{0.02\textwidth} 
  \begin{subfigure}[t]{0.3\textwidth}
    \centering
    \includegraphics[width=\textwidth]{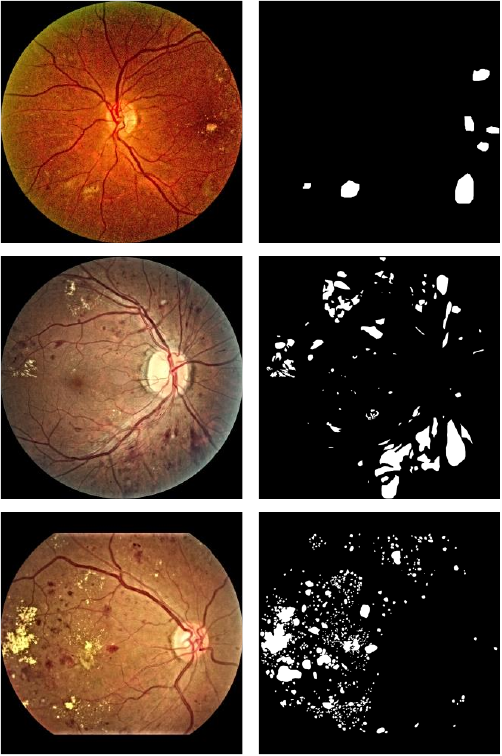}
    \caption{}
    \label{fig:example_4_b}
  \end{subfigure}
  \caption{Examples of the retina on the (a) EyePACS dataset: demonstrates retinal fundus images of varying qualities and characteristics, captured using various fundus cameras, such as Canon CR-2 and Topcon NW400, (b) FGADR dataset: the first column indicates the original images, and the second column indicates the lesion maps.}
  \label{fig:4}
\end{figure}

\begin{table}[t]
    \caption{Configuration of the EyePACS dataset.}
    \label{tab:1}
    \centering
    \renewcommand{\arraystretch}{1.5} 
    \setlength{\tabcolsep}{12pt} 
    \begin{tabular}{c|c|c|c|c|c}
    \specialrule{0.9pt}{0pt}{0pt}
    \multicolumn{2}{c|}{Train} & \multicolumn{2}{c|}{Validation} & \multicolumn{2}{c}{Test} \\ \hline
    0 & 1 & 0 & 1 & 0 & 1 \\ \hline
    28253 & 6873 & 8850 & 2056 & 34445 & 8225 \\ \hline
    \multicolumn{2}{c|}{35126} & \multicolumn{2}{c|}{10906} & \multicolumn{2}{c}{42670} \\ \hline
    \multicolumn{6}{c}{88702} \\ \specialrule{0.9pt}{0pt}{0pt}
    \end{tabular}
\end{table}

\noindent The EyePACS\cite{21} and FGADR\cite{22} datasets were used to validate the proposed model. 
The EyePACS dataset contains a large collection of image-level retinal fundus images of varying qualities and characteristics, captured using various fundus cameras.
Images in the EyePACS dataset range from 1024$\times$1024 to 2592$\times$1944 pixels, taken with devices like the Canon CR-2 and Topcon NW400. 
The FGADR dataset provides pixel-level high-resolution images (3072$\times$2048 pixels) and detailed ground truth lesion maps, essential for training the segmentation model. 
Both datasets offer a comprehensive set of images and labels necessary for robust training and evaluation. Out of the 88,702 images in the EyePACS dataset, 35,126 were used for training, 10,906 for validation, and 42,670 for testing. 
The EyePACS dataset categorizes diabetic retinopathy into five stages: no DR, mild DR, moderate DR, severe DR, and proliferative diabetic retinopathy (PDR). 
In this study, we follow the referable classification as in\cite{23}: no DR and mild DR are labeled as 0 (negative), and moderate DR, severe DR, and PDR are labeled as 1 (positive). Table~\ref{tab:1} shows the configuration of the EyePACS dataset. 
Fig.~\ref{fig:4} shows examples from the EyePACS dataset, highlighting the dataset's size and diversity.

\begin{figure}[t]
  \centering
  \includegraphics[height=3.5cm]{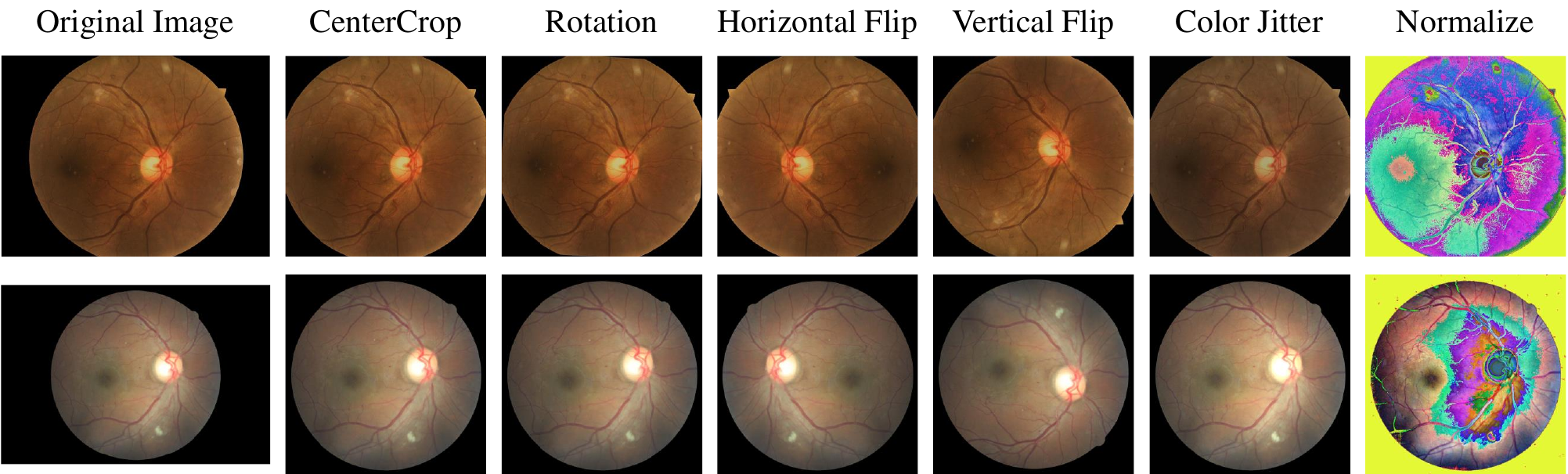}
  \caption{Examples of applying preprocessing and data augmentation on the EyePACS dataset. The first column shows the original images and the subsequent columns sequentially show the results of applying the preprocessing on the original image.}
  \label{fig:5}
\end{figure}

To ensure consistent performance, we applied several preprocessing steps. Initially, all images were resized to 640$\times$640 pixels to ensure uniformity. 
Center cropping removed unnecessary background information, focusing on the central region of the retina. 
Next, images were normalized using the mean and standard deviation values derived from the ImageNet dataset to properly adjust pixel intensity values. 
These preprocessing steps standardized the input data, allowing the model to concentrate on relevant features without being influenced by variations in image acquisition conditions. 
Also, various data augmentation techniques were employed to enhance the robustness and generalization capability of the model. 
These include random rotations, horizontal and vertical flips, and color jittering. Data augmentation exposes the model to a diverse range of image transformations during training, thereby mitigating overfitting. 
For instance, random rotations help the model recognize lesions regardless of their orientation, while color jittering improves its ability to detect features under varying lighting conditions. 
Elastic distortions and random cropping further augment variability. 
These augmentations are crucial in medical imaging, where variability in image acquisition significantly impacts model performance. 
Fig.~\ref{fig:5} illustrates examples of augmented images from the EyePACS dataset, showcasing the diversity introduced through these techniques. 
By augmenting the training data, the model's ability to generalize to new, unseen images is improved, ensuring robust performance in clinical settings.

We conducted extensive experiments using the EyePACS and FGADR datasets to evaluate the performance of our proposed method. 
The datasets were divided into training and validation sets, and the test sets. 
All experiments were executed on a system equipped with four NVIDIA RTX 4090 GPUs, each with 24GB of memory. 
The implementation was conducted using Python 3.8.11 and PyTorch 1.10.2. The Swin U-Net model was utilized for lesion segmentation, pre-trainedon ImageNet, and fine-tuned on the FGADR dataset. 
For the classification task, we employed the Swin-T model, incorporating the cross-attention mechanism. 
The models were trained using the SGD optimizer with a learning rate of 0.01, momentum of 0.9, and weight decay of 0.0001. 
The batch size was set to 16, and the models were trained for 50 epochs.

\subsection{Experiments and Results}
\noindent The model's performance was evaluated using multiple metrics. For the segmentation task, we used the Dice coefficient and Intersection over Union(IoU) to measure the accuracy of lesion segmentation. 
The Dice coefficient gauges the similarity between the predicted lesion map and the ground truth by calculating the overlap, ranging from 0 to 1, where 1 indicates perfect agreement. 
IoU measures the ratio of the intersection of the predicted and true lesion areas to their union, with a value closer to 1 indicating better segmentation performance. 
For the classification task, we employed True Positive Rate(TPR), True Negative Rate(TNR), Area Under the Curve(AUC), and overall accuracy(ACC). TPR, or sensitivity, measures the proportion of actual positives correctly identified by the model, while TNR, or specificity, measures the proportion of actual negatives correctly identified. 
AUC evaluates the trade-off between sensitivity and specificity across different thresholds, with higher values indicating better performance. 
ACC provides a general measure of the model's performance by calculating the ratio of correctly predicted instances to the total instances. 
These metrics provide a comprehensive evaluation of the model's ability to accurately classify referable DR and measure the accuracy of lesion segmentation, ensuring robust performance in clinical settings.

\begin{figure}[t]
  \centering
  \includegraphics[width=\linewidth, height=5cm]{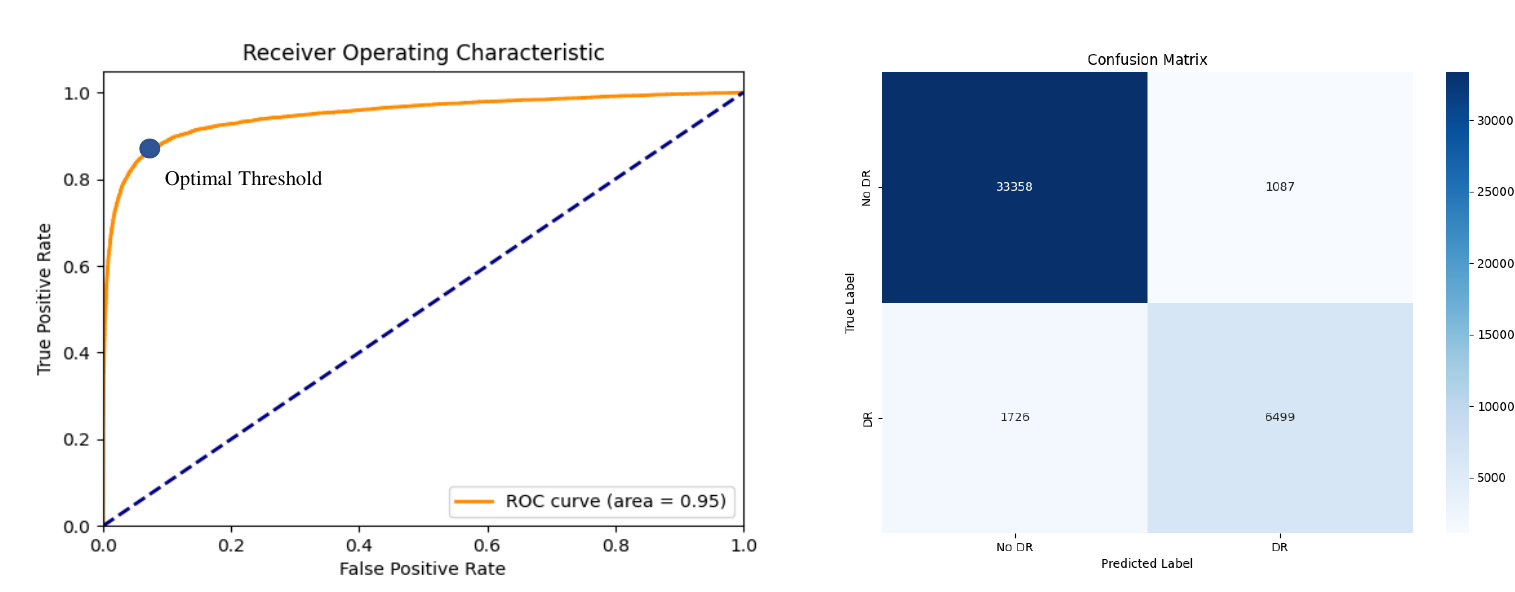}
  \caption{ROC curve and confusion matrix of the proposed model evaluated on the EyePACS dataset.}
  \label{fig:6}
\end{figure}

The proposed method achieved superior performance compared to existing state-of-the-art methods. 
In the segmentation task, the Swin U-Net model achieved a Dice coefficient of 0.89 and an IoU of 0.86 on the FGADR dataset. 
In the classification task, the Swin-T model, incorporating the cross-attention mechanism, achieved an AUC of 96.2\% and an overall accuracy of 94.6\% on the EyePACS dataset. 
Fig.~\ref{fig:6} provides a confusion matrix and an ROC curve illustrating the classification performance. 
To further enhance the performance of our classification model, we utilized the Receiver Operating Characteristic(ROC) curve to determine the optimal threshold for class distinction. 
The ROC curve plots the true positive rate(TPR) against the false positive rate(FPR) at various threshold settings, allowing us to identify the threshold that maximizes the trade-off between sensitivity and specificity. 
We identified the optimal threshold by finding the point on the ROC curve closest to the top-left corner (i.e., the point with the highest TPR and lowest FPR). 
This threshold was then applied to the classification model to make final predictions. 
Utilizing the optimal threshold derived from the ROC curve, we achieved a more balanced performance across various metrics, thereby enhancing the model's robustness and reliability. 
This step is crucial in clinical settings where the accurate classification of referable DR can significantly impact patient outcomes.

\begin{figure}[t]
  \centering
  \includegraphics[height=5cm]{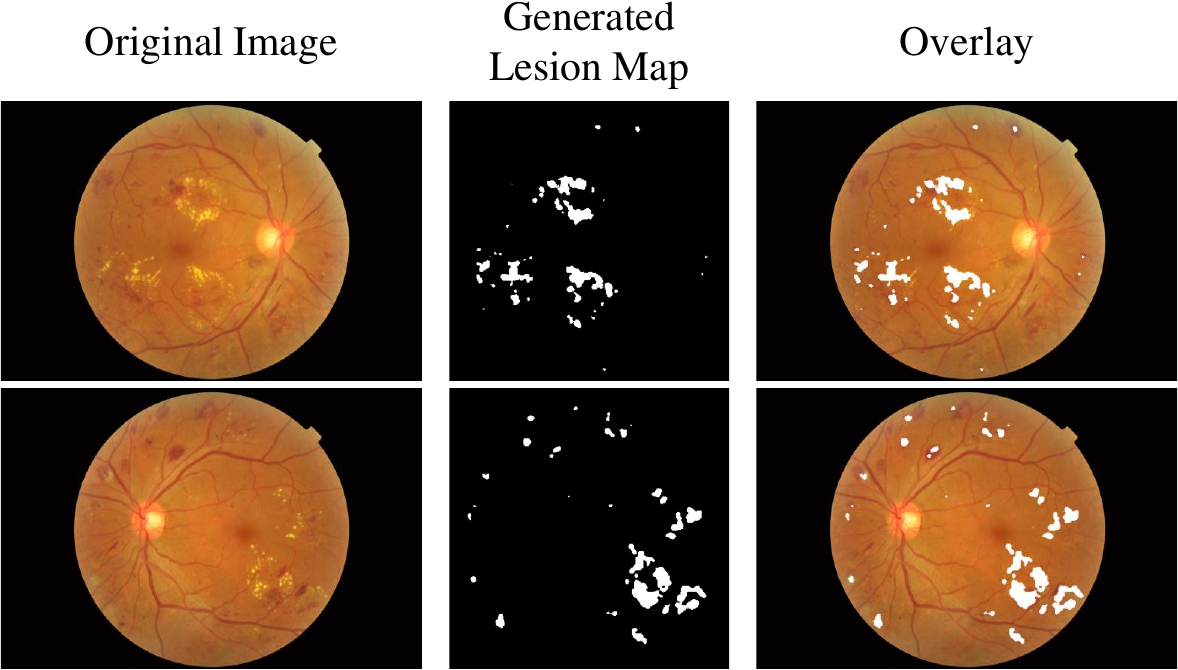}
  \caption{Examples of generated lesion map on EyePACS dataset.
  }
  \label{fig:7}
\end{figure}

\begin{table}[t]
  \caption{Performance comparisons with other studies on the EyePACS dataset.}
  \label{tab:2}
  \centering
  \begin{tabular}{p{7cm}cccc}
    \toprule
    Model & ACC & AUC & TPR & TNR \\
    \midrule
    ResNet50\cite{24} & 87.8 & 88.1 & 67.2 & 93.1 \\
    Swin-T(Tiny)\cite{25} & 91.6 & 91.7 & 81.3 & 92.2 \\
    Swin-S(Small)\cite{25} & 91.7 & 91.8 & 81.7 & 93.1 \\
    Swin-B(Base)\cite{25} & 91.7 & 91.7 & 81.8 & 93.4 \\
    Swin-L(Large)\cite{25} & 91.9 & 91.9 & 81.7 & 91.7 \\
    ConvNeXt\cite{26} & 87.5 & 85.1 & 76.4 & 81.6 \\
    CrossViT\cite{27} & 88.0 & 85.6 & 76.8 & 83.2 \\
    CoATNet\cite{28} & 81.4 & 79.4 & 72.6 & 77.3 \\
    Ensemble of Plain and Robust model\cite{06} & 90.2 & 91.0 & 79.2 & 88.1 \\
    Lesion-Aware Contrastive Learning\cite{20} & 84.6 & 83.8 & 73.4 & 85.2 \\
    \textbf{Ours (Proposed Method)} & \textbf{94.6} & \textbf{96.2} & \textbf{82.0} & \textbf{97.6} \\
    \bottomrule
  \end{tabular}
\end{table}

We deeply analyzed the performance of our model in classifying referable DR. 
The results revealed that the cross-attention mechanism substantially enhanced the model's capability to identify relevant features from both the original image and the lesion map.
Fig.~\ref{fig:7} shows the lesion map generated by Swin U-Net from the EyePACS dataset, demonstrating that the model effectively recognizes lesions. 
This enhancement was particularly prominent in cases with subtle lesion characteristics, where traditional methods faced challenges. 
The model's high AUC and accuracy indicate its robustness and reliability in clinical settings, where accurate and timely diagnosis is critical. 
Our method demonstrates significant potential for enhancing diagnostic accuracy and efficiency in detecting referable DR.

\begin{table}[t]
  \caption{Ablation study evaluated on the EyePACS dataset.}
  \label{tab:3}
  \centering
  \renewcommand{\arraystretch}{1} 
  \setlength{\tabcolsep}{6pt} 
  \begin{tabular}{cccc}
    \specialrule{0.9pt}{0pt}{0pt}
    \multirow{2}{*}{\begin{tabular}[c]{@{}c@{}}Cross Feature Fusion of Original Image \\ and Generated Lesion Map \end{tabular}} & \multirow{2}{*}{\begin{tabular}[c]{@{}c@{}}Transfer Learning \\ from Swin U-Net Encoder\end{tabular}} & \multirow{2}{*}{ACC} & \multirow{2}{*}{AUC} \\
     &  &  &  \\ \specialrule{0.3pt}{0pt}{0pt}
    - & - & 91.6 & 91.5 \\
    - & \checkmark & 93.4 & 95.0 \\
    \checkmark & - & 94.0 & 95.9 \\
    \checkmark & \checkmark & \textbf{94.6} & \textbf{96.2} \\ \specialrule{0.9pt}{0pt}{0pt}
  \end{tabular}
\end{table}

To evaluate the proposed method, we conducted a comparison with several state-of-the-art techniques, highlighting the strengths and weaknesses of each approach.
Table~\ref{tab:2} presents a comparative performance analysis between our model and various existing methods, including CNN-based models such as ResNet50, Swin Transformer models(Swin-T, Swin-S, Swin-B, Swin-L), and latest models such as ConvNeXt, CrossViT, and CoATNet.
Additionally, we compared our approach with existing methods\cite{20,06} for further analysis.
V. Boreiko \etal,\cite{06} combined robust and plain models in an ensemble to enhance robustness and interoperability. However, their approach excluded mild DR cases from the experiment, limiting the generalizability of their findings across all DR severity levels.
S. Cheng \etal,\cite{20} introduced lesion-aware contrastive learning to improve DR diagnosis accuracy, focusing on lesion-centric feature learning. However, their method required complex processes and a sophisticated setup to achieve high accuracy, posing challenges for practical implementation.
In contrast, our proposed method integrates cross-attention with the Swin-T model, demonstrating superior performance across all evaluated metrics under the same experimental conditions. Our method outperformed the state-of-the-art model by 4.4\% in terms of accuracy. By effectively combining features from the original images and generated lesion maps, the proposed method enhances accuracy, interpretability, and robustness while maintaining computational efficiency and scalability, making it highly suitable for clinical practice.

Table~\ref{tab:3} presents the results of an ablation study to evaluate the effectiveness of key components in our proposed method.
This study investigates the impact of cross-feature fusion between the generated lesion maps and the original images, as well as the effect of transfer learning using the Swin U-Net encoder.
By comparing performance with and without these components, we observed that the inclusion of both cross-feature fusion and transfer learning resulted in the highest performance.
The cross-attention mechanism, which integrates information from the original images and generated lesion maps, proved to be a critical factor in improving model accuracy, particularly by focusing on lesion-specific features.
Overall, the results of this ablation study validate the effectiveness of the proposed method in improving the accuracy and robustness of automated DR screening systems.

\section{Conclusion}
\noindent In this study, we have developed an advanced cross-learning method for the classification of referable DR using deep learning techniques. 
Our method leverages the Swin U-Net architecture to enhance classification accuracy by generating lesion maps and incorporating a cross-attention mechanism. 
By integrating the Swin-T and U-Net architectures for segmentation and combining them with an optimized thresholding approach, we have constructed a highly accurate model that can address the challenge of limited segmentation annotations. 
We leverage a relatively small amount of labeled data from the FGADR dataset to generate segmentation annotations for a large unlabeled dataset, namely EyePACS.
Experimental results on the FGADR and EyePACS datasets demonstrate superior performance compared to existing state-of-the-art methods, achieving AUC of 96.2\% and an overall accuracy of 94.6\%. 
Future research will include more diverse datasets to enhance generalizability, implement strategies to improve robustness, incorporate Grad-CAM for better interpretability, and investigate scalability to handle larger datasets and different medical imaging challenges. 
The findings from this study have significant clinical implications, enhancing the accuracy and efficiency of automated DR screening and improving patient outcomes by facilitating timely therapeutic interventions. 
This study lays a strong foundation for using advanced deep learning models in automated DR diagnosis, addressing dataset diversity and model interpretability to enhance performance and reliability in clinical applications further.

\begin{credits}

\subsubsection{\ackname} This work was supported in part by IITP grant funded by the Korean government (MSIT) under the ICT Creative Consilience program (RS-2020-II201821, 30\%), Development of Brain Disease (Stroke) Prediction Model based on Fundus Image Analysis (RS-2024-00459512, 30\%), Artificial Intelligence Graduate School Program (RS-2019-II190421, 20\%), and Artificial Intelligence Innovation Hub (RS-2021-II212068, 20\%).

\subsubsection{\discintname}
The authors have no competing interests to declare that are
relevant to the content of this article.
\end{credits}


%
%

\end{document}